\documentclass[a4paper]{jpconf}%
\usepackage{graphicx}
\usepackage{amsmath}
\usepackage{mathrsfs}
\usepackage{amsfonts}
\usepackage{dsfont}
\usepackage{feynmp}
\usepackage{caption}
\usepackage{amssymb}%
\begin{document}

\title{The eLSM at nonzero density}
\author{Francesco Giacosa}
\address{Institute of Physics, Jan Kochanowski University, ul. Swietokrzyska 15, 25-406 Kielce, Poland
and  Institute for Theoretical Physics, J. W. Goethe University,
Max-von-Laue-Str. 1, 60438 Frankfurt am Main, Germany} \ead{fgiacosa@ujk.edu.pl}

\begin{abstract}
The extended Linear Sigma Model (eLSM) is an effective model of QCD which
includes in the mesonic sector (pseudo)scalar and (axial-)vector quarkonia
mesons as well as one dilaton/glueball field and in the baryonic sector the
nucleon doublet and its chiral partner in the mirror assignment. The chiral
partner of the pion turns out to be the resonance $f_{0}(1370),$ which is then
predominantly a quarkonium state. As a consequence, $f_{0}(500)$ is
predominately not a quarkonium state but a four-quark object and is at first
not part of the model. Yet, $f_{0}(500)$ is important in the baryonic sector
and affects nuclear matter saturation, the high-density behavior, and
nucleon-nucleon scattering. In these proceedings, we show how to enlarge the
two-flavour version of the eLSM in order to include the four-quark field
$f_{0}(500)$ in a chiral invariant manner. We then discuss homogeneous and
inhomogeneous chiral restoration in a dense medium.

\end{abstract}

\section{Introduction}

Linear Sigma Models (LSMs) are effective models of QCD which contains hadrons
(mesons and baryons) as degrees of freedom and which are based on the linear
realization of chiral symmetry \cite{lee}. As a consequence of spontaneous
breaking of chiral symmetry, in these models the pions emerge as
quasi-Goldstone bosons (a relatively small mass is present because of explicit
breaking of chiral symmetry). The chiral partner of the pion, denoted as
$\sigma_{N},$ is also an explicit d.o.f. of such models. As various studies
show, this meson corresponds to the scalar resonance $f_{0}(1370)$: this state
is then predominantly a quark-antiquark state. Hence, the lightest scalar
state listed in the PDG \cite{pdg}, the resonance $f_{0}(500),$ must be
something else: its substructure corresponds to a four-quark state, either as
diquark-antidiquark or pion-pion enhancement (see e.g. the recent review on
the subject \cite{sigmareview}). As such, this resonance should not be (at
first) part of a LSM. Yet, as we shall describe later on, $f_{0}(500)$ is
important at nonzero density since it describes a necessary middle-range
attraction between nucleons (see also Ref. \cite{machleidtrev}).

Extensions of LSMs toward the inclusion of (axial-)vector degrees of freedom
were performed in Ref. \cite{ko}. Quite recently, a complete as possible LSM,
called extended Linear Sigma Model (eLSM), was developed. The eLSM contains
from the very beginning (axial-)vector fields and embodies both chiral
symmetry and dilatation invariance. Spontaneous and explicit breaking of the
former as well as anomalous breaking of the latter are present. As a
consequence of chiral symmetry and the dilation invariance, the eLSM contains
a \textit{finite} number of terms. The eLSM was first presented for $N_{f}=2$
in Refs. \cite{elsmnf2}, then it was enlarged to $N_{f}=3$ in\ Refs.
\cite{dick} (this is the first version of a chiral model with $N_{f}=3$
containing (axial-)vector d.o.f.), and lately it was also studied for charmed
mesons ($N_{f}=4$, Ref. \cite{walaa}).

In the baryonic sector, the eLSM was investigated for \ $N_{f}=2$ in\ Refs.
\cite{susannaold} (a first step toward the eLSM at $N_{f}=3$ was performed in
Ref. \cite{lisa}). In the eLSM, the mirror assignment, which allows for
chirally invariant mass term, is used \cite{detar,zische}. As mentioned in
Ref. \cite{susannaold}, a four-quark field $\chi$ corresponding to
$f_{0}(500)$ can be coupled to the eLSM in a chirally invariant way, see Sec.
2. The eLSM with $f_{0}(500)$ has been investigated at nonzero density
in\ Ref. \cite{susagiu}, where the chiral phase transition has been
investigated, and later on in\ Ref. \cite{achim2}, in which inhomogeneous
condensation has been studied; these results are here summarized in\ Sec. 3.
The role of the additional, non-conventional meson $f_{0}(500)$ turns out to
be important: it makes a description of the properties of nuclear matter
possible (both saturation and compressibility are in agreement with data) and
it strongly affects the properties of nuclear matter at high density. Quite
interestingly, the resonance $f_{0}(500)$ was recently investigated in\ Ref.
\cite{kt} in the framework of nucleon-nucleon scattering: also in this case,
its presence is necessary for a correct description of data.

\section{The eLSM}

\subsection{eLSM without $f_{0}(500)$}

We briefly present the eLSM for $N_{f}=2$. (Pseudo)scalar mesons are contained
in $\Phi=(\sigma_{N}+i\eta_{N})t^{0}+(\vec{a}_{0}+i\vec{\pi})\cdot\vec{t}$,
where $t^{0}=1_{2}/2$, $\vec{t}=\vec{\sigma}/2,$ $\sigma_{i}$ are the Pauli
matrices. In Table 1 we report the identification of the fields with
resonances of the PDG \cite{pdg}. [Note: $\eta_{N}\equiv\left(  u\bar{u}%
+d\bar{d}\right)  /\sqrt{2}$ reads $\eta_{N}=\cos\varphi_{P}\eta-\sin
\varphi_{P}\eta^{\prime}$ with $\varphi_{P}\approx-44^{\circ}$ \cite{dick}]
Because of spontaneous symmetry breaking, $\sigma_{N}$ condenses: $\sigma
_{N}\rightarrow\sigma_{N}+\phi,$ where $\phi$ is the chiral condensate. Under
$U(2)_{R}\times U(2)_{L}$ chiral transformations: $\Phi\rightarrow U_{L}\Phi
U_{R}^{\dagger}.$

The left-handed and right-handed fields $L^{\mu}$ and $R^{\mu}$ contain the
vector states $\omega^{\mu}$ and $\vec{\rho}^{\mu}$ and the axial--vector
states $f_{1}^{\mu}$ and $\vec{a}_{1}^{\mu}$: $L^{\mu}=(\omega^{\mu}%
+f_{1}^{\mu})t^{0}+(\vec{\rho}^{\mu}+\vec{a}_{1}^{\mu})\cdot\vec{t}$ $,$
$R^{\mu}=(\omega^{\mu}-f_{1}^{\mu})t^{0}+(\vec{\rho}^{\mu}-\vec{a}_{1}^{\mu
})\cdot\vec{t}$, see Table 1. Under chiral transformations: $L^{\mu
}\rightarrow U_{L}\Phi U_{L}^{\dagger}$ and $R^{\mu}\rightarrow U_{R}\Phi
U_{R}^{\dagger}$.

\begin{center}
\textbf{Table 1}: Correspondence of eLSM $\bar{q}q$ fields to\ PDG \cite{pdg}.%

\begin{tabular}
[c]{|c|c|c|c|c|c|}\hline
Field & PDG & Quark content & $I$ & $J^{PC}$ & Mass (GeV)\\\hline
$\pi^{+},\pi^{-},\pi^{0}$ & $\pi$ & $u\bar{d},d\bar{u},\frac{u\bar{u}-d\bar
{d}}{\sqrt{2}}$ & $1$ & $0^{-+}$ & $0.13957$\\\hline
$\eta$ & $\eta(547)$ & $\frac{u\bar{u}+d\bar{d}}{\sqrt{2}}\cos\varphi
_{P}-s\bar{s}\sin\varphi_{P}$ & $0$ & $0^{-+}$ & $0.54786$\\\hline
$\eta^{\prime}$ & $\eta^{\prime}(958)$ & $\frac{u\bar{u}+d\bar{d}}{\sqrt{2}%
}\sin\varphi_{P}+s\bar{s}\cos\varphi_{P}$ & $0$ & $0^{-+}$ & $0.95778$\\\hline
$a_{0}^{+},a_{0}^{-},a_{0}^{0}$ & $a_{0}(1450)$ & $u\bar{d},d\bar{u}%
,\frac{u\bar{u}-d\bar{d}}{\sqrt{2}}$ & $1$ & $0^{++}$ & $1.474$\\\hline
$\sigma_{N}$ & $f_{0}(1370)$ & $\frac{u\bar{u}+d\bar{d}}{\sqrt{2}}$ & $0$ &
$0^{++}$ & $1.350$\\\hline
$\rho^{+},\rho^{-},\rho^{0}$ & $\rho(770)$ & $u\bar{d},d\bar{u},\frac{u\bar
{u}-d\bar{d}}{\sqrt{2}}$ & $1$ & $1^{--}$ & $0.77526$\\\hline
$\omega_{N}$ & $\omega(782)$ & $\frac{u\bar{u}+d\bar{d}}{\sqrt{2}}$ & $0$ &
$1^{--}$ & $0.78265$\\\hline
$a_{1}^{+},a_{1}^{-},a_{1}^{0}$ & $a_{1}(1230)$ & $u\bar{d},d\bar{u}%
,\frac{u\bar{u}-d\bar{d}}{\sqrt{2}}$ & $1$ & $1^{++}$ & $1.230$\\\hline
$f_{1,N}$ & $f_{1}(1285)$ & $\frac{u\bar{u}+d\bar{d}}{\sqrt{2}}$ & $0$ &
$1^{++}$ & $1.2819$\\\hline
\end{tabular}

\end{center}

The mesonic part of the Lagrangian reads%

\begin{align}
\mathcal{L}_{eLSM}^{meson}  &  =Tr\left[  (D^{\mu}\Phi)^{\dagger}(D^{\mu}%
\Phi)\right]  -\mu_{0}^{2}Tr[\Phi^{\dagger}\Phi]-\lambda_{2}Tr[(\Phi^{\dagger
}\Phi)^{2}]-\frac{1}{4}Tr[(L^{\mu\nu})^{2}+(R^{\mu\nu})^{2}]\nonumber\\
&  +\frac{m_{1}^{2}}{2}Tr\left[  L^{\mu\ \!\!2}\!+\!R^{\mu\ \!\!2}\right]
+Tr[H(\Phi+\Phi^{\dagger})]+h_{2}\mathrm{Tr}\left[  \Phi^{\dag}L_{\mu}L^{\mu
}\Phi+\Phi R_{\mu}R^{\mu}\Phi\right]  +2h_{3}\mathrm{Tr}\left[  \Phi R_{\mu
}\Phi^{\dag}L^{\mu}\right]  \text{ }...\text{,} \label{lmes}%
\end{align}
where dots refer to large-$N_{c}$ suppressed terms (including the chiral
anomaly). In Refs. \cite{elsmnf2,dick} it was shown that, thanks to the
inclusion of (axial-)vector d.o.f., the eLSM provides a good description of
meson phenomenology. An interesting consequence is that the quark-antiquark
field $\sigma_{N},$ which represents chiral partner of the pion, is associated
to $f_{0}(1370)$, in agreement with previous phenomenological studies
\cite{close}. Hence, $f_{0}(500)$ must be something else \cite{sigmareview}.

We now turn to the baryonic sector. For $N_{f}=2$ one starts from two nucleon
fields $\Psi_{1}$ and $\Psi_{2}$ with opposite parity which transform
mirror-like under chiral transformations: $\Psi_{1,R(L)}\rightarrow
U(2)_{R(L)}\Psi_{1,R(L)},$ $\Psi_{2,R(L)}\rightarrow U(2)_{L(R)}\Psi
_{2,R(L)}.$ The eLSM Lagrangian is \cite{susannaold}:%

\begin{align}
\mathcal{L}_{eLSM}^{baryons}  &  =\bar{\Psi}_{1L}i\gamma_{\mu}{D}_{1L}^{\mu
}\Psi_{1L}+\bar{\Psi}_{1R}i\gamma_{\mu}{D}_{1R}^{\mu}\Psi_{1R}+\bar{\Psi}%
_{2L}i\gamma_{\mu}{D}_{2R}^{\mu}\Psi_{2L}+\bar{\Psi}_{2R}i\gamma_{\mu}{D}%
_{2L}^{\mu}\Psi_{2R}\nonumber\\
&  -\hat{g}_{1}(\bar{\Psi}_{1L}\Phi\Psi_{1R}+\bar{\Psi}_{1R}\Phi^{\dagger}%
\Psi_{1L})-\hat{g}_{2}(\bar{\Psi}_{2L}\Phi^{\dagger}\Psi_{2R}+\bar{\Psi}%
_{2R}\Phi\Psi_{2L})\nonumber\\
&  -m_{0}(\bar{\Psi}_{1L}\Psi_{2R}-\bar{\Psi}_{1R}\Psi_{2L}-\bar{\Psi}%
_{2L}\Psi_{1R}+\bar{\Psi}_{2R}\Psi_{1L})\;, \label{lbar}%
\end{align}
where $D_{1(2)R(L)}^{\mu}=\partial^{\mu}-ic_{1(2)}R(L)^{\mu}$. The fields
$\Psi_{1}$ and $\Psi_{2}$ mix due to the $m_{0}$-term and are related to the
physical states $N$ and its chiral partner $N^{\ast}$ via:
\begin{equation}
\Psi_{1}=\frac{1}{\sqrt{2\cosh\delta}}\left(  Ne^{\delta/2}+\gamma_{5}N^{\ast
}e^{-\delta/2}\right)  \text{ },\text{ }\Psi_{2}=\frac{1}{\sqrt{2\cosh\delta}%
}\left(  \gamma_{5}Ne^{-\delta/2}-N^{\ast}e^{\delta/2}\right)  \text{ },
\end{equation}
where $\cosh\delta=\frac{m_{N}+m_{N^{\ast}}}{2m_{0}}.$ The field $N$
corresponds to the nucleon $N(939)$ and $N^{\ast}$ to $N(1535)$ or $N(1650)$.
For the purposes of the present work, the assignment of $N^{\ast}$ is
marginal, see however \cite{susannaold,lisa}. The parameter $m_{0}$ represents
a chirally invariant mass, which was first discussed in Ref.\ \cite{detar} and
further investigated in Refs.\ \cite{susannaold,zische}. The masses of the
nucleon $N$ and its chiral partner $N^{\ast}$ are given by:
\begin{equation}
m_{N,N^{\ast}}=\sqrt{m_{0}^{2}+\frac{(\hat{g_{1}}+\hat{g_{2}})^{2}}{16}%
\phi^{2}}\pm\frac{1}{4}(\hat{g_{1}}-\hat{g_{2}})\phi\text{ }.
\end{equation}
In the limit $m_{0}\rightarrow0$ one obtains the result $m_{N}=\hat{g_{1}}%
\phi/2$, i.e., the nucleon mass is solely generated by the chiral condensate
[as in the original Linear Sigma Model \cite{lee}].The parameters of the model
were determined in\ Ref. \cite{susannaold}, to which we refer for details.

\subsection{Inclusion of $f_{0}(500)$ in the eLSM}

We now introduce $\chi\equiv f_{0}(500)$ with quantum numbers $I(J^{PC}%
)=0(0^{++})$ and mass $m_{\chi}=(0.475\pm0.25)$ GeV \cite{pdg} into the eLSM.
This state is regarded as an admixture of $\pi\pi$ and $[u,d][\bar{u},\bar
{d}]$ configurations. For $N_{f}=2$ it is a singlet under chiral
transformations ($\chi\rightarrow\chi$). The coupling of $\chi$ to baryons is
obtained by modifying the $m_{0}$-term as:%
\begin{equation}
m_{0}(\bar{\Psi}_{1L}\Psi_{2R}-\bar{\Psi}_{1R}\Psi_{2L}-\bar{\Psi}_{2L}%
\Psi_{1R}+\bar{\Psi}_{2R}\Psi_{1L})\rightarrow a\chi(\bar{\Psi}_{1L}\Psi
_{2R}-\bar{\Psi}_{1R}\Psi_{2L}-\bar{\Psi}_{2L}\Psi_{1R}+\bar{\Psi}_{2R}%
\Psi_{1L})\text{ ,}%
\end{equation}
where $a$ is now a dimensionless constant. Then, the mass parameter $m_{0}$
emerges as a condensation of the four-quark field $\chi$: $m_{0}=a\chi_{0}$ .
In the mesonic sector, one has too add
\begin{equation}
\mathcal{L}_{eLSM}^{meson}\rightarrow\mathcal{L}_{eLSM}^{meson}+\frac{1}%
{2}\left(  \left(  \partial_{\mu}\chi\right)  ^{2}-m_{\chi}^{2}\chi
^{2}\right)  +g_{\chi\Phi}\chi Tr[\Phi^{\dagger}\Phi]+g_{\chi\Phi}\chi
Tr[L^{\mu\ \!\!2}\!+\!R^{\mu\ \!\!2}]+...\text{ ,}%
\end{equation}
where dots refer to large-$N_{c}$ suppressed terms. As a consequence, the
condensate $\chi_{0}$ takes the form $\chi_{0}=g_{\chi\Phi}\phi^{2}/m_{\chi
}^{2}$.

\section{Results}

\subsection{Homogenous condensation}

First, we study nuclear matter at nonzero density under the assumption that
the condensates are homogenous. Two scalar fields condense: $\left\langle
\sigma_{N}\right\rangle =\phi(\mu)$ and $\left\langle \chi\right\rangle
=\bar{\chi}(\mu)$, where $\mu$ is the nuclear chemical potential. The results
are obtained by minimizing -at a given $\mu$- the thermodynamical potential
$\Omega$ w.r.t. $\phi,$ $\bar{\chi},$ as well as $\left\langle \omega
^{0}\right\rangle $. The vacuum's relation $\chi_{0}=g_{\chi\Phi}\phi
^{2}/m_{\chi}^{2}$ holds approximately also at nonzero density: $\phi$ slowly
decreases as function of $\mu$ together with $\bar{\chi}.$ Then, at a critical
$\mu_{c}^{\text{hom}}\sim1$ GeV (corresponding to a density $\rho/\rho_{0}%
\sim2.7,$ where $\rho_{0}$ is the nuclear matter saturation density) a
first-order phase transition takes place: $\phi$ and $\bar{\chi}$ drop to to
very small (but nonzero) values. Chiral symmetry is restored (see Fig. 1, left
panel). In terms of density, there is a long mixed phase between $2.7\rho_{0}%
$-$10\rho_{0}.$ The compressibility $K$ lies in the range $200$-$250$ MeV and
is therefore in good agreement with the experiment ($200$-$300$ MeV). The mass
of the nucleon and that of the partner drop to almost zero in the chirally
restored phase. For details, see Ref. \cite{susagiu}.%

\begin{figure}
[ptb]
\begin{center}
\includegraphics[
height=2.1482in,
width=5.5659in
]%
{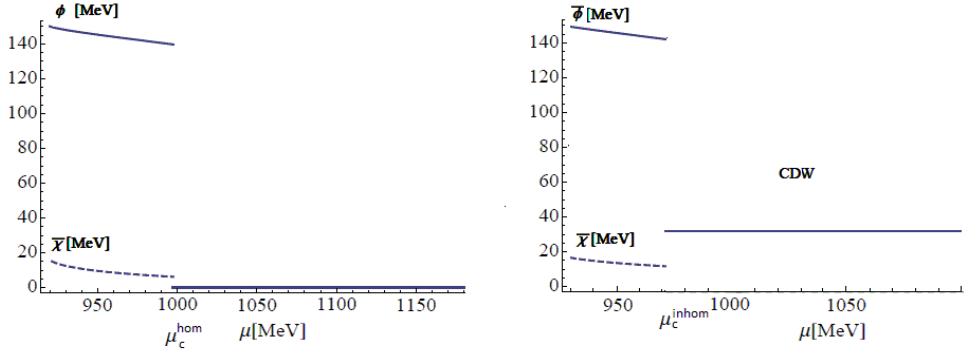}%
\caption{Left: homogenous condensation.\ The condensates $\phi$ and $\bar
{\chi}$ drop to almost zero at $\mu_{c}^{\text{hom}}.$ Right: inhomogeneous
condensation. At $\mu_{c}^{\text{inhom}}$ a transition to the inhomogeneous
condensation of Eq. (\ref{cs}) takes place.\ This configuration is more
favorable than the homogenous case of the left panel.}%
\end{center}
\end{figure}

\subsection{Inhomogeneous condensation}

The previous results were obtained under the assumption that the $\phi$ and
$\bar{\chi}$ are homogenous, i.e. are not space-dependent. Yet, various
studies have shown (see Ref. \cite{carignano} and refs. therein) that an
inhomogeneous condensation can be favoured. In Ref. \cite{achim2} the
so-called chiral-spiral \cite{cs} was investigated:
\begin{equation}
\phi(\mu,z)=\bar{\phi}\cos(2fz)\text{, }\left\langle \pi^{3\equiv
0}\right\rangle =\bar{\phi}\sin(2fz) \label{cs}%
\end{equation}
with $\bar{\phi}\equiv\bar{\phi}(\mu)$ and $f\equiv f(\mu)$. The homogenous
case corresponds to $f=0.$ For $f\neq0$ one has a condensation of the neutral
pion field, which corresponds to a spontaneous breaking of parity at nonzero
density. At a given $\mu,$ the thermodynamical potential is now minimized for
$\bar{\phi},$ $f,$ $\bar{\chi},$ and $\left\langle \omega^{0}\right\rangle $.
The results show that below a certain critical chemical potential $\mu
_{c}^{\text{inhom}}\lesssim1$ GeV one has $f=0$: homogenous condensation is
realized, just as before. However, at $\mu_{c}^{\text{inhom}}$ a phase
transition takes place: $f$ jumps to a finite value (of about $400\ $MeV,
which then slowly increases with $\mu)$, and $\bar{\phi}$ to a lower but
finite value, see Fig. 2. Interestingly, for a given parameter set, it turns
out that $\mu_{c}^{\text{inhom}}<\mu_{c}^{\text{hom}}$: the homogenous chiral
phase transition does not occur, but is only a local minimum. Chiral symmetry
is only partially restored.

\section{Conclusions}

The resonance $f_{0}(500)$ is the lightest scalar listed in the PDG and hence
is potentially interesting in hadron phenomenology. Its role has to be
investigated case by case. While its condensate may be relevant at nonzero
temperature \cite{achim1}, its effect on thermal models turns out to be
negligible because of a very subtle and interesting cancellation with the
repulsive isotensor channel: for practical purposes, $f_{0}(500)$ can be
neglected in thermal models of heavy ion collisions \cite{nosigma}.

At nonzero density, $f_{0}(500)$ plays indeed a significant role because it
mediates a sizable attraction between nucleons: in these proceedings, we have
incorporated $f_{0}(500)$ into the eLSM and reviewed the properties at finite
chemical potential. We have shown that inhomogeneous condensation of the
chiral-spiral type is favored w.r.t. homogeneous one. In the future, one
should go beyond the chiral-spiral Ansatz and test arbitrary types of
inhomogeneous condensation by using the numerical procedure put forward in
Ref. \cite{achimlast}.

\section*{Acknowledgments}

The author thanks all the members of the chiral group in Frankfurt for the
many valuable cooperations and discussions.

\end{document}